\documentclass{icrc29}
\usepackage{graphicx,amssymb,amsmath,times}
\begin{document}
%Title of paper
\title[The FLASH Thick Target Experiment]{The FLASH Thick Target Experiment: 
 Direct Measurement of Air Fluorescence Yield in Electromagnetic Showers}
\author [J. Belz] {J. Belz$^a$ for the FLASH Collaboration \\
        (a) University of Montana, Missoula, Montana USA and 
            University of Utah, Salt Lake City, Utah USA}
\presenter{Presenter: J. Belz (belz@cosmic.utah.edu), \  
           usa-belz-J-abs3-he15-oral}

\maketitle

\begin{abstract}

A key assumption in the reconstruction of extensive air showers 
using the air fluorescence technique is the hypothesis that 
fluorescence is proportional to energy deposition at all depths
in the shower. This ansatz, along with the supposition that
particle distribution and energy loss can be well modeled by 
modern shower simulation software, must be thoroughly verified 
in order to validate the air fluorescence technique. We report
here the results of the first direct measurement of air 
fluorescence yield as a function of shower depth, as performed
in the thick-target phase of the FLASH (FLuorescence in Air from 
SHowers) experimental program at the SLAC Final-Focus Test Beam
facility. We compare observed fluorescence light yields as a 
function of shower depth to concurrently measured charged particle 
yields, to the predictions of the EGS and GEANT software packages, 
and to empirical energy-loss models. We also examine the extent to which 
the relative yield versus shower depth is independent of wavelength
within the fluorescence spectrum. 

\end{abstract}

\section{Introduction}

The cosmic ray spectrum above 10$^{19}$~eV (1.6 Joules per particle) is not well understood from either the theoretical or experimental point of view. Mechanisms that could lead to these energies have been postulated, either by acceleration from very energetic sources~\cite{gaiser95,mannheim95,waxman97} or by decay of primordial super heavy particles~\cite{cirkel98,battacharjee92,protheroe96,birkel98,battacharjee00}, but strong supporting evidence remains to be reported. At the same time, the spectrum reported by the AGASA detector~\cite{takeda99}, an array of scintillators covering 100 sq km at ground level, is both more intense and extends to higher energy than that of the atmospheric fluorescence detector, HiRes~\cite{abassi04}. At least the former result appears to violate the cutoff in the spectrum expected from interactions with the cosmic microwave background, the GZK effect~\cite{greisen66,kuzmin66} just below 10$^{20}$~eV. Further experiments are needed to clarify the situation, and to enhance the presently very limited statistics. There are several under consideration, in planning or under construction~\cite{abraham04,fukushima02,scarsi02,linsley99}. All of these include at least a fluorescence measurement system.
     
The FLASH (FLuorescence in Air SHowers) program at the Stanford Linear Accelerator Center (SLAC) is intended to provide an experimental basis for the use of atmospheric fluorescence in imaging showers from ultra-high energy cosmic rays (UHECR). This paper reports on the first study of the longitudinal profile of air fluorescence light in electromagnetic showers. 

Other aspects of the fluorescence technique that are under experimental testing by various groups are the absolute yield of light in the relevant wavelength band, and its spectrum, as a function of atmospheric pressure. This is done at several fixed electron beam energies~\cite{kakimoto96,nagano04,belz02}. Of course, energy loss to the gas atoms is a function of the energy of the charged shower particles, changing rapidly below the minimum that occurs at about 1.5 MeV. For this reason, the work discussed here makes use of actual showers to examine the precision with which simulations of shower development and energy loss, and actual ionization measurements, agree with the profile as measured using the fluorescent light.

\section{The Experiment: Fluorescence Yield and Beam Longitudinal Profile} 

The work described here is a study of the longitudinal shower profile in the beam at the SLAC Final Focus Test Beam (FFTB) facility, using electrons delivered in 5 ps long pulses of a few $\times 10^7$ electrons per pulse, at 28.5 GeV. We note that the energy of the pulse initiating the electromagnetic shower is therefore of order $10^{18}$~eV. As a practical and economic way of simulating the effect of air, we have chosen to use a commercially available alumina ceramic. The material, delivered in brick form, is Al$_2$O$_3$ with 10\% SiO$_2$. The measured mean density was 3.51 g cm$^{-3}$. The radiation length, 28 g cm$^{-2}$, is just 24\% shorter than that of air, and the critical energy, below which ionization energy loss dominates, is 54 MeV, compared with 87 MeV for air. 

\begin{figure}[h]
\begin{center}
  \includegraphics[width=0.75\columnwidth]{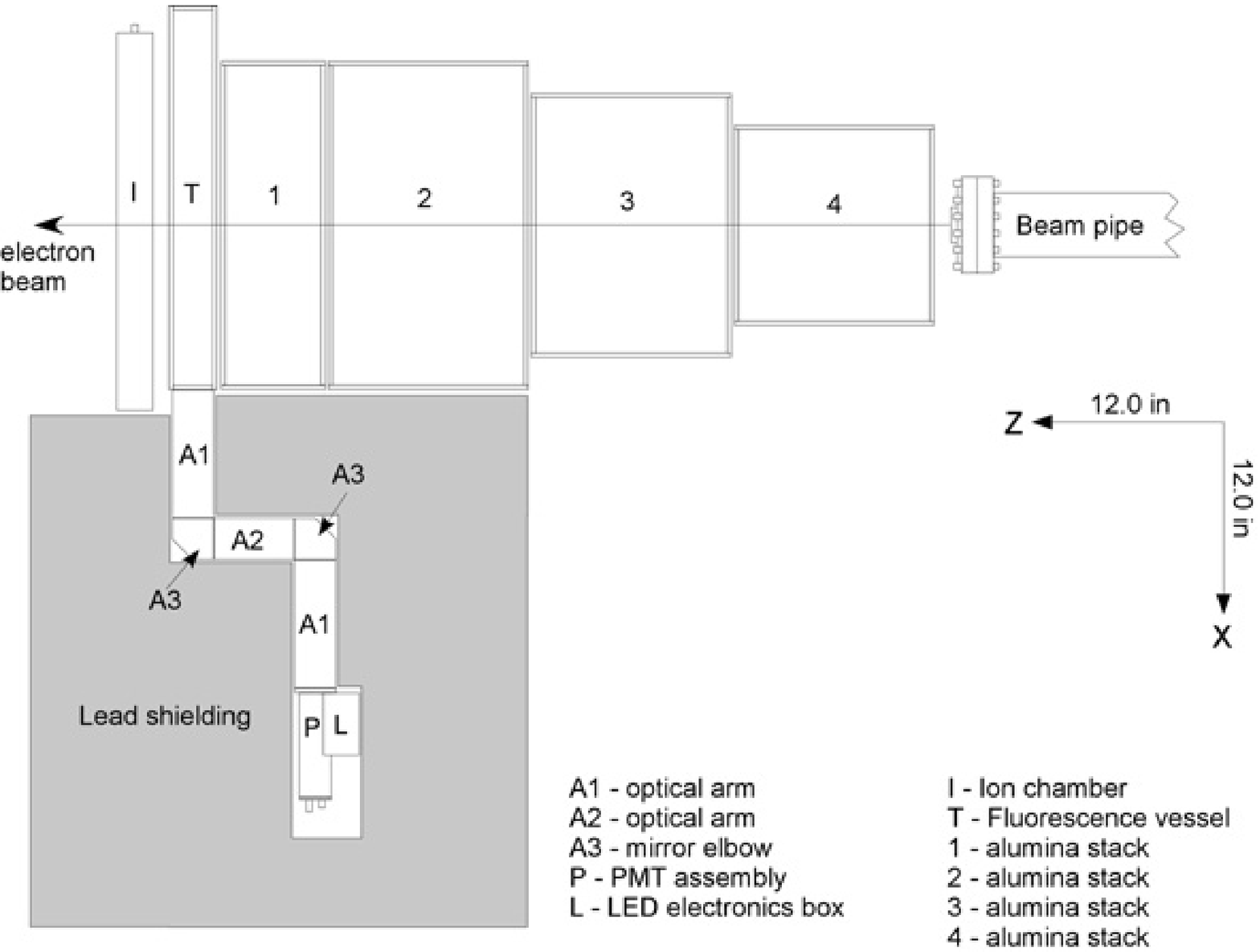}
  \caption{Schematic view of detector apparatus.}
  \label{layout}
\end{center}
\end{figure}

A schematic view of the apparatus is shown in Figure~\ref{layout}. It was installed in a gap in the electron beam vacuum pipe. The electron beam exited through a thin window, The alumina was contained in a line of four aluminum boxes that could remotely and independently be moved on or off the beam line. The downstream block was approximately 2 radiation lengths (15cm) thick, by 50 cm wide, and the air fluorescence detector was placed immediately behind it. Each of the upstream blocks was 4 radiation lengths thick. This arrangement permitted thicknesses of approximately 0 to 14 radiation lengths in 2 radiation length steps to be selected.In this way the longitudinal profile of an electromagnetic shower could be developed. The shower particles leaving the alumina immediately entered the detector volume, where they caused a flash of fluorescence in the layer of atmospheric pressure air. The detector was in the form of a flat rectangular aluminum box, its air space 4.0 cm thick along the beam direction, and with vertical dimension, 50 cm, matching the alumina. In order to allow the electron beam to pass through with minimal scattering for tests and set-up, the aluminum walls were thinned to 25 microns for a diameter of 7.8 cm about the beam. 

Some of the light traveled towards a vertical row of photomultiplier tubes mounted on one side. It was necessary to take steps to suppress the accidental collection of the forward  going Cherenkov light from the air as well as fluorescence light scattered from the walls. After wall scattering, these would have an uncertain spectrum and collection efficiency. The suppression was done in the standard way, using a set of 1cm wide vertical baffles on the front and back walls, and all surfaces, except mirrors and photomultiplier tube (PMT) apertures, were covered with black flock material.

In order to shield the PMTs from ionizing radiation from the showers, the light path was built with two 90 degree reflections, as seen in Fig~\ref{layout}. After these, at a horizontal path length of 91 cm from the beam line, there were apertures for the PMTs. This design allowed for a wall of lead to protect the PMTs from the radiation emitted from the side walls of the alumina, or from scattering sources nearby. The minimum thickness of the lead was 25 radiation lengths.

The ion chamber was designed for the high radiation and ionization levels, and wide dynamic range, encountered after the shower media. It used 11 active gaps, nominally 0.9 mm thick, with plates based on printed circuit board covering the 50 cm square active width of the air fluorescence chamber. The gas was helium at 1 atmosphere, and the applied voltage, 140 V/mm, was chosen to maximize the clearing field and electrode charge without leading to gas gain. All anodes were connected electrically, as were all cathodes. Their signals were read out without amplification.

\begin{figure}[h]
\begin{center}
  \includegraphics[width=0.75\columnwidth]{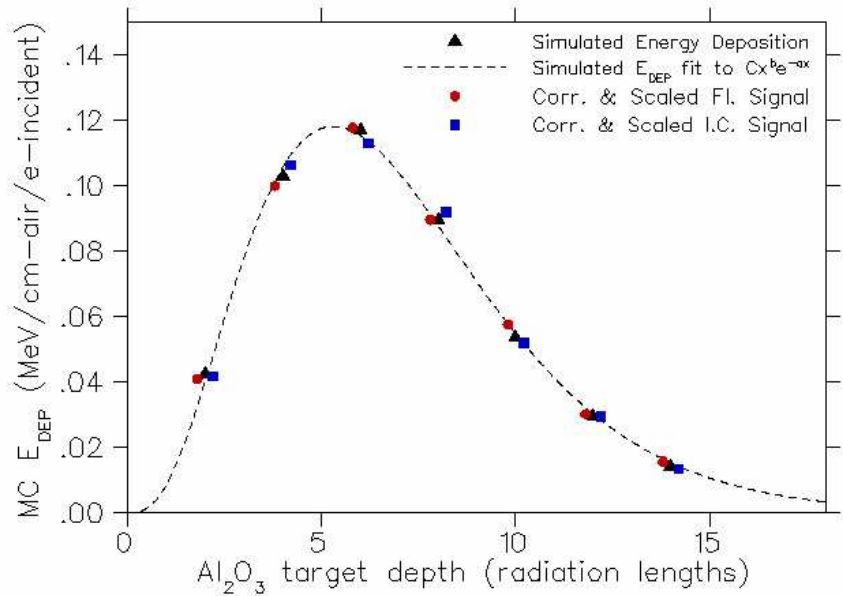}
  \caption{Detector response as a function of shower depth. The simulated 
  energy deposition (triangles), measured fluorescence signal (circles) 
  and ion chamber signal (squares) are superimposed on a fit to an empirical
  energy loss model~\cite{eidelman04}.}
  \label{dedx}
\end{center}
\end{figure}

For each thickness of alumina, the fluorescence vessel and ion chamber signals were plotted pulse by pulse against the toroid signals. The slopes of the resultant correlation plots were interpreted as being proportional to the air fluorescence yield and beam longitudinal profile, respectively. 

The resulting profiles versus radiation length are shown in Figure~\ref{dedx}. The data taken with the compact alumina arrangements (0, 2, 6, 10 , 14 rad. lengths) and the sets with the air gap (4, 8, 12 rad. lengths) are both shown. Geometrical correction factors have been applied to the ``air gap'' sets to account for particle shadowing by the downstream alumina block. For comparison, the results of a GEANT~3.2~\cite{geant} simulation are also shown, along with a curve corresponding to an empirical energy loss model~\cite{eidelman04}. The agreement is quite adequate for the requirements of this study.

\section{Conclusions}
 
The measurements reported here confirm the validity of the technique of imaging and measuring electromagnetic showers in the atmosphere using fluorescent emission from the air. Further details of the results of the studies described in this paper will be presented at ICRC2005.

\section{Acknowledgements}

We are indebted to the SLAC accelerator operations staff for 
their expertise in meeting the unusual beam requirements, and
to personnel of the Experimental Facilities Department for very 
professional assistance in preparation and installation of 
equipment. We also gratefully acknowledge the many contributions
from the technical staffs of our home institutions. This work was 
supported in part by the U.S. Department of Energy under contract
number DE-AC02-76SF00515 as well as by the National Science 
Foundation under awards NSF PHY-0245428, PHY-0305516, PHY-0307098 
and PHY-0400053.


\begin{thebibliography}{99}


\bibitem{gaiser95} T.K. Gaiser, F. Halzen and T.Stanev, 
                   Phys. Rep. 258 (1995) 173 

\bibitem{mannheim95} K. Mannheim, Astropart. Phys. 3 (1995) 295

\bibitem{waxman97} E. Waxmann and J.N. Bahcall, Phys. Rev. Lett. 
                   78 (1997) 2292.

\bibitem{cirkel98} M. Cirkel and S. Sarkar, Astropart. Phys 9, (1998) 297; 

\bibitem{battacharjee92} P. Battacharjee, C.T. Hill and D.N. Schramm 
                         Phys. Rev. Lett. 69, (1992) 567

\bibitem{protheroe96} R.J. Protheroe and T. Stanev, Phys. Rev. Lett. 
                      77 (1996) 3708; 

\bibitem{birkel98} M. Birkel and S. Sarkar, Astropart. Phys. 9 (1998) 297

\bibitem{battacharjee00} P. Battacharjee and G. Sigl, 
                         Phys. Rep. 327 (2000) 109.

\bibitem{takeda99} M. Takeda et al., Astrophys. J. 522 (1999) 225.

\bibitem{abassi04} R. Abassi et al., Phys. Rev. Letters 92 (2004) 151101.

\bibitem{greisen66} K. Greisen, Phys. Rev. Letters 16 (1966) 748

\bibitem{kuzmin66} V.A. Kuzmin, G.T. Zatsepin, Pisma Zh.
                   Eksp. Teor. Fiz. 4 (1966) 114, JETP Letters 4,78.

\bibitem{abraham04} J. Abraham et al., Nucl. Instr. Meth. A 523 (2004) 50.

\bibitem{fukushima02} M. Fukushima, Institute for Cosmic Ray Research Mid-Term 
            (2004-2009) Maintenance Plan Proposal Book ``Cosmic Ray 
            Telescope Project'', Tokyo Univ., (2002).

\bibitem{scarsi02} L. Scarsi et al., Proc. 27th Intl. Cosmic Ray Conf.,
                   Copernicus Gesellschaft, Hamburg, (2002) 175, 

\bibitem{linsley99} J, Linsley, Proc., 26th Intl. Cosmic Ray Conf., 
                    Univ. Utah, Salt Lake City, (1999), Vol. 2, 423.

\bibitem{eidelman04} For a summary and review, seel S. Eidelman et al.,
                     Phys. Letters B592 (2004), 242.

\bibitem{kakimoto96} F. Kakimoto et al., Nucl. Instr. Meth. A 372 (1996) 527; 

\bibitem{nagano04} M. Nagano et al., Astropart. Phys. 22 (2004) 235; 

\bibitem{belz02} J. Belz et al., SLAC Experimental Proposal E-165 (2002).

\bibitem{geant} CERN Applications Software Group, Geneva, Switzerland (1993).

\end{thebibliography}
\end{document}